\documentclass[
reprint,
nofootinbib,
preprintnumbers,
amsmath,amssymb,
aps,
prd
]{revtex4-2}

\usepackage{dcolumn}
\usepackage{bm}
\usepackage{appendix}
\usepackage{amsmath}
\usepackage{slashed}
\usepackage{multirow}
\usepackage{color}
\usepackage{mathtools}
\usepackage{graphicx}
\usepackage{braket}
\usepackage{setspace}
\usepackage{graphicx}
\usepackage{geometry}
\usepackage{hyperref}
\usepackage{starfont}
\usepackage{amssymb}
\usepackage{bbold}
\usepackage[shortlabels]{enumitem}
\usepackage{framed}
\usepackage{empheq}
\usepackage{amsmath}
\usepackage{tikz-feynman}
\usepackage{simpler-wick}
\usepackage{wasysym}
\usepackage{mhchem}
\usepackage{extarrows}
\usepackage{aas_macros}

\DeclarePairedDelimiter\abs{\lvert}{\rvert}
\makeatletter
\let\oldabs\abs
\def\abs{\@ifstar{\oldabs}{\oldabs*}}
\makeatother

\def\bea{\begin{eqnarray}}
\def\eea{\end{eqnarray}}

\def\Mp{M_{\rm pl}}

\def\GeV{\,{\rm GeV}}
\def\MeV{\,{\rm MeV}}

\def\eV{\,{\rm eV}}

\begin{document}
\preprint{DESY-22-088}

\title{Oscillations of atomic energy levels induced by QCD axion dark matter}

\author{Hyungjin Kim}
\email{hyungjin.kim@desy.de}
\affiliation{%
Deutsches Elektronen-Synchrotron DESY, Notkestr.\,85, 22607 Hamburg, Germany}

\author{Gilad Perez}
\email{gilad.perez@weizmann.ac.il}
\affiliation{%
Department of Particle Physics and Astrophysics, Weizmann Institute of Science, Rehovot, Israel 7610001}

\begin{abstract}
Axion-gluon interaction induces quadratic couplings between the axion and the matter fields. We find that, if the axion is an ultralight dark matter field, it induces small oscillations of the mass of the hadrons as well as other nuclear quantities. As a result, atomic energy levels oscillate. We use currently available atomic spectroscopy data to constrain such axion-gluon coupling. We also project the sensitivities of future experiments, such as ones using molecular and nuclear clock transitions. We show that current and near-future experiments constrain a finely-tuned parameter space of axion models. These can compete with or dominate the already-existing constraints from oscillating neutron electric dipole moment and supernova bound, in addition to those expected from near future magnetometer-based experiments. We also briefly discuss the reach of accelerometers and interferometers.
\end{abstract}

\maketitle

We consider axion models, consisting of a pseudo-scalar field $a$ with the following coupling to the gluon field strength,
\bea
{\cal L} = \frac{g_s^2}{32\pi^2} \frac{a}{f} G_{\mu\nu}^a \widetilde{G}^{a\mu\nu},
\label{phiGGdual}
\eea
where $f$ is an axion decay constant, $g_s$ is a strong coupling and $\widetilde{G}^{a\mu\nu}$ is the dual gluon field strength.
Below the QCD scale, the above axion-gluon interaction induces axion coupling to the hadronic states. 
The pion mass depends on the axion field as
$$
m_\pi^2(\theta) = B \sqrt{m_u^2 + m_d^2 + 2 m_u m_d \cos\theta}\,.
$$
Here $\theta = a /f$ and $B = - \langle \bar q q \rangle_0 / f_\pi^2 $ with a pion decay constant $f_\pi \simeq 93\MeV$. 
The resulting axion potential can be described by $V(\theta) = -m_\pi^2(\theta) f_\pi^2$ to leading order~\cite{DiVecchia:1980yfw}. 
Due to the $\theta$-dependent potential, the axion relaxes to the CP conserving vacuum, thereby solving the strong CP problem dynamically~\cite{Peccei:1977hh, Peccei:1977ur, Weinberg:1977ma, Wilczek:1977pj, Kim:1979if, Shifman:1979if, Zhitnitsky:1980tq, Dine:1981rt}. 

Axion oscillation around its minimum may comprise dark matter (DM) in the present universe~\cite{Preskill:1982cy, Abbott:1982af, Dine:1982ah}.
If so, the pion mass develops a subdominant oscillatory component, given by
\bea
\frac{\delta m_\pi^2 }{m_\pi^2} = - \frac{m_u m_d}{2(m_u + m_d)^2} \theta^2\,. 
\label{delta_pi}
\eea
Other nuclear quantities such as hadron masses and magnetic moments consist of similar oscillating contributions, all induced by the effective quadratic coupling between the axion and the matter fields.
It results in a corresponding time-variation of the atomic energy levels, which can be probed by monitoring transition frequencies of stable frequency standards.
This method was suggested by Arvanitaki et al.~\cite{Arvanitaki:2014faa} for dilaton/scalar DM searches (or a relaxion DM~\cite{Banerjee:2018xmn}), where the DM field naturally couples to the field strength of the strong and electromagnetic interactions as well as fermion masses.
It was also suggested that interferometry~\cite{Stadnik:2014tta}, as well as accelerometry~\cite{Graham:2015ifn}, can be used to probe the time-variation of fundamental constants induced by DM. 
Various experimental techniques have been used to search for such scalar-SM interactions~\cite{VanTilburg:2015oza, Branca:2016rez, Hees:2016gop, Aharony:2019iad, Antypas:2019qji, Grote:2019uvn, Savalle:2020vgz, Kennedy:2020bac, Campbell:2020fvq, Vermeulen:2021epa, Aiello:2021wlp, Oswald:2021vtc, Tretiak:2022ndx}. 
See Refs.~\cite{Safronova:2017xyt, Antypas:2022asj} for recent reviews.

\begin{figure}[h]
\centering
\includegraphics[width=0.48\textwidth]{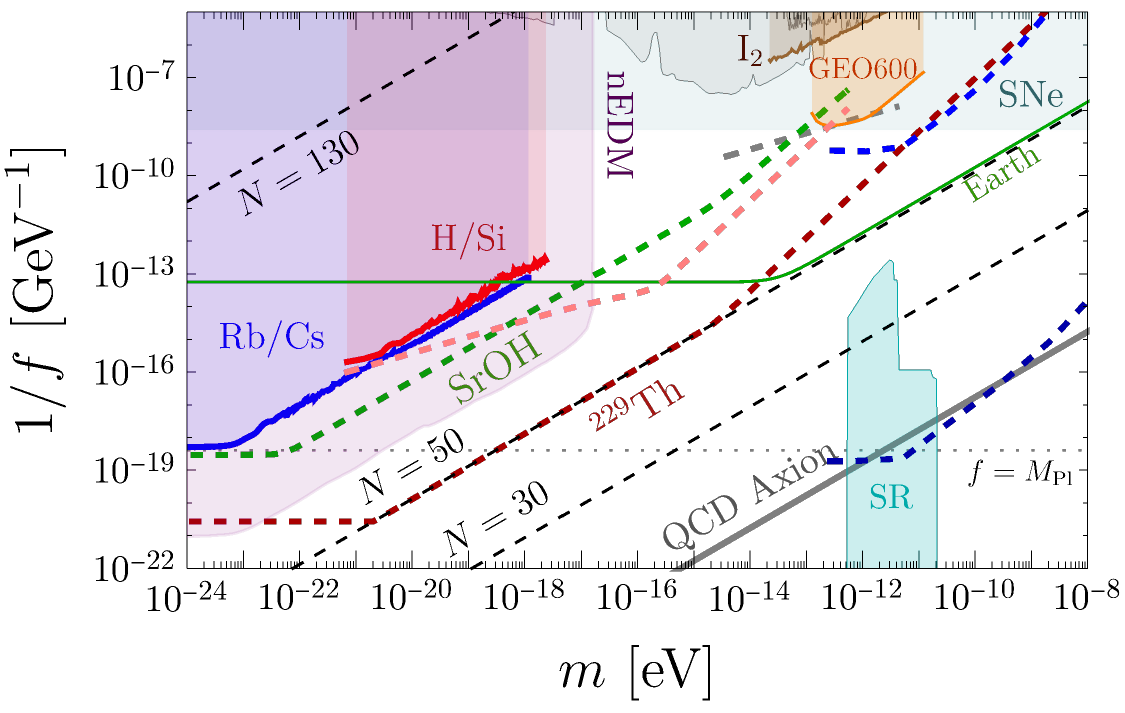}
\caption{%
 Constraints and future projections on the axion-gluon coupling are summarized as follows: Rb/Cs clock comparison (blue)~\cite{Hees:2016gop}, H/Si comparison (red)~\cite{Kennedy:2020bac}, Iodine molecular spectroscopy (brown)~\cite{Oswald:2021vtc}, GEO 600 gravitational wave detector (orange)~\cite{Grote:2019uvn}, torsion pendulum (pink dashed)~\cite{Graham:2015ifn}, $^{229}$Th nuclear isomer transition (red dashed)~\cite{Arvanitaki:2014faa, Banerjee:2020kww}, and strontium monohydroxide (green dashed)~\cite{Kozyryev:2018pcp}.
The gray dotted line is $f= \Mp$.
The diagonal grey line is allowed parameter space for the QCD axion, $m^2 f^2 \sim m_\pi^2 f_\pi^2$. 
Other bounds, such as oscillating neutron EDM (purple)~\cite{Abel:2017rtm}, supernova 1987A~\cite{Raffelt:2006cw} (light cyan), co-magnetometer and NASDUCK~\cite{Bloch:2019lcy, Bloch:2021vnn} (gray), and axion superradiance~\cite{Arvanitaki:2014wva} (cyan), are also included for the comparison.
Projections of axion-nucleon interaction searches, such as CASPEr-electric (blue dashed)~\cite{Aybas:2021nvn} and NASDUCK (gray dashed)~\cite{Bloch:2021vnn}, are also included.
Spectroscopy bounds above the green solid line must be taken carefully as the axion could develop a static profile around the earth~\cite{Hook:2017psm}. 
If such a static profile exists, it affects the propagation of DM axion, but this parameter space is already excluded by static neutron EDM experiments. 
The black dashed lines denote the required number of copies of the standard model to achieve the parameter space above the QCD axion line in the model of Hook~\cite{Hook:2018jle}.
See the main text for details.}
\label{fig:summary}
\end{figure}

The goal of this work is to assess the possibility of whether the axion-gluon coupling can be probed by the same method, i.e. by monitoring atomic energy levels of stable frequency standards. 
We claim that the same principle can be applied to probe the coupling~\eqref{phiGGdual}.
We show the current constraints and projections of future experiments as well as other constraints in Figure~\ref{fig:summary}. 
We explain the main idea below.

For the purpose of demonstration, we consider the ground state hyperfine transition in the hydrogen atom. 
The hyperfine structure arises due to the interaction between the electron magnetic moment and the magnetic field generated by the proton magnetic moment. 
The transition frequency of the ground state hydrogen hyperfine structure is
$$
f_{H} = \frac{2}{3\pi} \frac{g_p m_e^2 \alpha^4}{m_p} 
\simeq 1420\,{\rm MHz},
$$
where $g_p = 5.586$ is the proton $g$-factor.
In the presence of axion dark matter and the axion-gluon coupling, the proton $g$-factor and proton mass develop a small oscillating component, and so does transition frequency $f_H$. 
The fractional variation of hyperfine transition frequency can be written as
\bea
\frac{\delta f_{H}}{f_{H}} &=& 
\frac{\delta g_p}{g_p} - \frac{\delta m_p}{m_p}
\nonumber\\
&=& 
\left[ 
\frac{\partial \ln g_p}{\partial \ln m_\pi^2}
- \frac{\partial \ln m_p}{\partial \ln m_\pi^2}
\right] \frac{\delta m_\pi^2}{m_\pi^2}
\nonumber\\
&\simeq& 10^{-15} \times \frac{ \cos(2mt) }{m_{15}^2 f_{10}^2} 
\label{hfs}
\eea
where we have defined $m_{15} = m /10^{-15}\eV$ and $f_{10} = f / 10^{10}\GeV$, and used $\partial \ln g_p / \partial \ln m_\pi^2 \simeq -0.17$ and $\partial \ln m_p / \partial \ln m_\pi^2 \simeq 0.06$.
For now, we take $m$ and $f$ as independent parameters to investigate the reach of spectroscopy experiments for the axion-gluon coupling search. 
Axion DM background does not change the fine structure constant and electron mass to the leading order, so the variation of those quantities is ignored.
The dependence of $g_p$ and $m_p$ on the pion mass is computed by using chiral perturbation theory at the chiral order ${\cal O}(p^3)$~\cite{Gasser:1987rb, Bernard:1992qa, Scherer:2012xha, Hoferichter:2015tha} and compared with lattice computations~\cite{Aoki:2021kgd, Djukanovic:2021cgp}. 
See Appendix~\ref{app} for details. 
We have used $\theta^2 (t) = (\rho_{\rm DM} / m^2 f^2) [ 1 + \cos(2mt) ]$ with $\rho_{\rm DM} \simeq 0.4\GeV/{\rm cm}^3$. 
A constant offset is ignored as it is unobservable. 
Equation~\eqref{hfs} suggests that the axion-gluon coupling strength might be probed by looking for a harmonic signal in $\delta f_H / f_H$ at the frequency $\omega = 2m$. 

The above discussion is more than an academic exercise. 
A recent experiment performed by Kennedy et al~\cite{Kennedy:2020bac} monitored hydrogen maser frequency ($f_H$) together with silicon optical cavity resonance frequency ($f_{\rm Si}$) to probe scalar DM interactions to electromagnetic field strength and electron mass.  
Since the silicon optical cavity resonance frequency has a rather weak dependence on proton mass, the fractional variation of frequencies is dominated by that of hydrogen maser,
$$
\frac{\delta (f_H/f_{\rm Si})}{(f_H/f_{\rm Si})} 
\simeq \frac{\delta f_H}{f_H}.
$$
Claimed short-term stability of transition frequency is $\sim 3\times 10^{-13}/\sqrt{\rm Hz}$. 
Using Eq.~\eqref{hfs} and 33 days of experimental results obtained in Ref.~\cite{Kennedy:2020bac}, we place a constraint on axion-gluon coupling, shown as a red line in Figure~\ref{fig:summary}. 

Hydrogen maser is one example of many frequency standards based on hyperfine structure. 
An earlier attempt to probe scalar DM based on hyperfine transitions was made by Hees et al~\cite{Hees:2016gop}, where they used measurement of rubidium ($^{87}$Rb) and cesium ($^{133}$Cs) hyperfine transitions. 
For the hyperfine structure of heavier atoms, the parametric dependence of transition frequency is similar~\cite{Safronova:2017xyt},\footnote{In this work, we take $m_p = m_n = m_N$. While the mass difference $m_n - m_p$ also depends on $\theta$-angle, it only provides a subleading correction.}
\bea
f \propto g m_e^2 \alpha^4 /m_p,
\eea
but the $g$-factor is replaced by that of the nucleus.
The nuclear $g$-factor can be written as a function of nucleon $g$-factor and the spin expectation value of valence and core nucleons.
Using the result of Ref.~\cite{Flambaum:2006ip} together with the nucleon $g$-factor computed in the chiral perturbation theory, we find
\bea
\frac{\partial \ln g}{\partial \ln m_\pi^2} =
\begin{cases}
 -0.024 & \textrm{$^{87}$Rb},
\\
+ 0.011 & \textrm{$^{133}$Cs},
\end{cases}
\eea 
See Appendix~\ref{app} for details. 
The fractional frequency variation is therefore
\bea
\frac{\delta( f_A/f_B)}{(f_A / f_B)} 
&\simeq& -0.04 \frac{\delta m_\pi^2}{m_\pi^2}
\simeq - 10^{-16}  \frac{ \cos(2mt) }{m_{15}^2 f_{10}^2} 
\eea
where $A = \,^{87}{\rm Rb}$ and $B = \, ^{133}{\rm Cs}$. 
Using the experimental result of Rb/Cs fountain clock~\cite{Hees:2016gop}, we obtain a constraint on the axion-gluon coupling constant, which is shown as a blue line in Figure~\ref{fig:summary}. 
It is similar to the constraint from the H/Si comparison test, but Rb/Cs constraint extends to a much lower mass range due to its long experimental time scale.

We have only considered hyperfine transitions so far.
In principle, any stable frequency standards can be used for axion DM search as long as the transition frequency depends on $g$-factor and/or nucleon mass. 
Another example is a vibrational molecular excitation. 
Since the vibrational energy level for a diatomic molecule depends on $f_{\rm vib} \propto \mu^{-1/2}$ with the reduced mass $\mu = m_{A_1}  m_{A_2} / ( m_{A_1} + m_{A_2} )$, we find 
\bea
\frac{ \delta f_{\rm vib} }{  f_{\rm vib} }
= - \frac{1}{2} \frac{ \delta m_p }{ m_p }
\simeq  - 10^{-16} \times \frac{ \cos(2mt) }{m_{15}^2 f_{10}^2} .
\eea
A recent experiment performed by Oswald et al~\cite{Oswald:2021vtc} used molecular transitions in molecular iodine ($I_2$) to probe the variation of fundamental constants.
We use their result to place a constraint on axion-gluon coupling, which is shown as brown in the summary figure. 
The constraint is relevant for the relatively high mass end of the shown parameter space.

Better sensitivity to the variation of nucleon mass can be achieved by considering nearly degenerate vibrational energy levels in polyatomic molecules. 
It was shown in Ref.~\cite{Kozyryev:2018pcp} that a transition between nearly degenerate rovibrational energy levels in strontium monohydroxide (SrOH) could yield $\delta f / f = K \delta m_p / m_p$ with $K \sim {\cal O}(10^2)$. 
The projected sensitivity from SrOH is presented as a green dashed line in Figure~\ref{fig:summary}. 

As a final example, we consider a nuclear transition. 
Refs.~\cite{Campbell:2012zzb, 2003EL.....61..181P} suggested that an excitation in Thorium could be used as a frequency standard with extraordinary accuracy due to its anomalously low excitation energy and long lifetime. 
The latest measurement of such nuclear clock transition frequency is $\omega =8.3\eV$~\cite{Seiferle:2019fbe}. 
This low-lying excitation is due to the cancellation of binding energy from strong and electromagnetic interaction.
It suggests that even a small variation either in nuclear binding energy or electromagnetic binding energy will result in rather a large variation in transition frequency. 
The variation of transition frequency as a function of hadron mass is estimated in Refs.~\cite{Flambaum:2007mj, Flambaum:2008hu}. 
Using their result, one finds
\bea
\frac{\delta f_{\rm Th} }{f_{\rm Th}} \simeq  
(2 \times 10^5)  \frac{\delta m_\pi^2}{m_\pi^2}. 
\eea
This large sensitivity coefficient is due to the cancellation between nuclear and electromagnetic binding energy.
Assuming that noise is dominated by quantum projection noise of nuclear clock and that $t_{\rm int} = 10^{6}\, {\rm sec}$ of an integration time~\cite{Banerjee:2020kww}, we obtain the reach of nuclear clock transition for axion-gluon coupling in Figure~\ref{fig:summary} (red dashed).

Let us now discuss limitations of the analysis as well as other existing constraints for axion-gluon coupling in a similar mass range.

We assume that the axion comprises entire DM and that its field value on the surface of the earth is given by the local DM density, i.e.  $\theta(t) = (\sqrt{2\rho_{\rm DM}}/ m f) \cos(mt)$. 
The gravitational potential of the sun might modify the field value on the surface of the earth but such effect is generally an order of $(v_{\rm esc}/v_{\rm dm})^2 \sim 10^{-2}$ so that it can be ignored at the level of precision of the current analysis~\cite{Kim:2021yyo}. 

A more concerning effect is the finite density effect due to the sun or the earth.
As is already shown in Figure~\ref{fig:summary}, current constraints and future projections are above the QCD axion line. 
This range of parameter space is fine-tuned parameter space for minimal QCD axion models.  
For this parameter space, it was shown by Hook and Huang~\cite{Hook:2017psm} that the axion potential inside the sun or the earth could have a minimum at $\theta \neq 0$, developing static axion profile $\theta_s(r) \propto e^{-m_a r} /r$ (see also~\cite{Balkin:2020dsr,Budnik:2020nwz,Balkin:2021wea}). 
The green line in the figure corresponds to the parameter space where the earth sources the static axion profile. 
Such profile acts as an attractive potential for dark matter axion field, and hence the field value on the surface of the earth could be enhanced compared to its local value. 
While we acknowledge that this effect is likely to happen for a large part of the parameter space shown in the figure, we do not include this effect in our analysis since such parameter space is already constrained by experimental limits on neutron electric dipole moment.

We have argued that spectroscopy searches can constrain or probe the parameter space well above the QCD axion line, i.e. fine-tuned parameter space.
It is worth noting that such parameter space can be achieved in alternative QCD axion models in a natural way. 
Using a large discrete symmetry and $N$-copy of the standard model, it was shown that the parameter space above the QCD axion line can be obtained in a technically natural way~\cite{Hook:2018jle, DiLuzio:2021pxd, DiLuzio:2021gos}.
More specifically, from $m^2 f^2 \simeq N^{3/2} (m_u/m_d)^N m_\pi^2 f_\pi^2$~\cite{DiLuzio:2021pxd}, we find $N\sim 50\times\log\left[ \left(m_a/10^{-16}{\,\rm eV} \right)\left(f/10^{16}{\,\rm GeV}\right)\right]$.
See the black dashed lines in the figure.

Other atomic spectroscopy, interferometry, and accelerometry experiments can also be used for searches for axion-gluon interaction. 
Examples include frequency comparison between mechanical oscillators~\cite{Campbell:2020fvq}, unequal arm-length interferometry~\cite{Savalle:2020vgz}, interferometric gravitational wave detectors~\cite{Grote:2019uvn, Vermeulen:2021epa}, and torsion pendulum experiments~\cite{Graham:2015ifn}. 
All of these experiments depend on the change of nucleon mass to some degree.
Especially, the gradient of atomic mass is $\nabla M_A \propto \nabla \theta^2 \neq 0 $ and this leads to a periodic force in test masses, $\vec F_A = -\nabla M_A$, in interferometric gravitational wave experiments as well as in torsion pendulum experiments, which could induce observable signals. 
As an example, we show the constraint from the GEO~600 gravitational wave detector (orange) and a projection of the future torsion pendulum experiment (pink dashed) in the figure.
For the torsion pendulum experiment, we use the benchmark ``upgrade'' in Ref.~\cite{Graham:2015ifn}.

We do not consider frequency standards based on the electronic transition in this work.
This is because electronic transition frequencies scale as $f \propto M \alpha^{\xi}$ with the reduced mass $M = m_e M_N / (m_e + M_N)$ so that the effect of axion DM in electronic transition is always suppressed by $\propto m_e / M_N$ with nucleus mass $M_N$. 
It would still be interesting as a future direction to investigate the reach of optical atomic clocks for axion-gluon coupling search since uncertainties in optical clocks are much smaller than in clocks based on hyperfine structures. 

We compare spectroscopic experiments with other axion searches for axion-nucleon interactions. 
The axion-gluon coupling~(\ref{phiGGdual}) inevitably induces nucleon electric dipole moment (EDM), $d_n \simeq 2.4\times 10^{-16} \bar\theta \,{\rm e\cdot cm}$~\cite{Pospelov:1999mv}. 
The axion DM background leads to oscillating neutron EDM.
By measuring spin precession frequencies of neutron and mercury, Abel et al placed a strong constraint on axion-neutron coupling below $m \lesssim 10^{-17}\eV$~\cite{Abel:2017rtm}, which is shown as the purple region in the figure. 
A similar constraint was obtained from the EDM measurement of HfF$^+$~\cite{Roussy:2020ily}. 

Magnetometers provide another way to measure axion-nucleon coupling. 
The same axion-gluon coupling (\ref{phiGGdual}) induces axial vector coupling of axion to nucleon, ${\cal L} \supset C_\psi (\partial_\mu a /2f) \bar \psi \gamma^\mu \gamma_5 \psi$ with $C_p = - 0.47$ and $C_n = - 0.02$~\cite{GrillidiCortona:2015jxo}.
In the nonrelativistic limit, the way that the gradient of axion couples to nucleon is similar to the way that magnetic field couples to nucleon spin. 
Using magnetometers, Bloch et al placed constraints on axion-neutron axial-vector coupling for a wide range of axion mass, $10^{-24}\eV \lesssim m \lesssim 10^{-13}\eV$~\cite{Bloch:2019lcy, Bloch:2021vnn}. 
The constraint on $f$ (gray shaded) is similar to the one from Iodine molecular spectroscopy, while future projection (gray dashed) could dominate the supernova bound. 
In addition, the axion-nucleon coupling can be searched with nuclear magnetic resonance (NMR)~\cite{Graham:2013gfa, Budker:2013hfa}.
The current constraint using NMR technique is $f \gtrsim \GeV$ at $m\simeq 10^{-7}\eV$~\cite{Aybas:2021nvn}, while future projection with a much larger sample size (dark blue dashed) might reach to the QCD axion parameter space.

It is interesting to compare the reach of spectroscopic experiments with astrophysical and cosmological constraints. 
Axion-nucleon couplings provide an additional energy-loss channel for stellar objects.
Considering the observation of neutrino flux from supernova SN1987A, the axion-nucleon coupling is constrained as $f\gtrsim 4\times 10^8\GeV$~\cite{Raffelt:2006cw} (see also alternative claim by Bar et al~\cite{Bar:2019ifz}).
Neutron star provides additional constraints on axion-nucleon coupling~\cite{Iwamoto:1984ir, Leinson:2014ioa, Sedrakian:2015krq, Hamaguchi:2018oqw, Beznogov:2018fda, Leinson:2021ety, Buschmann:2021juv}. 
A recent study on the cooling of old neutron stars provides $ f \gtrsim 3 \times 10^{8}\GeV$~\cite{Buschmann:2021juv}, which is comparable to the supernova bound.

Another interesting bound, which is not shown in  Figure~\ref{fig:summary}, is the constraint from big bang nucleosynthesis (BBN)~\cite{Blum:2014vsa}. 
Blum et al took the cosmic dark matter density in the present universe, extrapolated the field value to the value at the time of BBN assuming the standard cosmological evolution of axion, $\theta(t_{\rm BBN}) \simeq \theta(t_0) \min[ (a_0/a_{\rm BBN})^{3/2}, (a_0 / a_{m})^{3/2}]$ with $a_m$ being scale factor at $3H = m$, and computed the modification of the helium 4 mass fraction due to the neutron/proton mass difference arising from axion field~\cite{Blum:2014vsa}. 
This bound constrains a large part of parameter space for the mass range shown in the figure.
The BBN bound might be avoided if, for some reason, the extrapolation fails and $\theta(t_{\rm BBN}) \lesssim 1$.
The deviation of proton-neutron mass difference from its standard value becomes quickly negligible for moderately small $\theta$~\cite{Lee:2020tmi}.

In summary, we have shown that the axion-gluon coupling induces an oscillation in nuclear quantities in the presence of axion dark matter and that it can be probed by atomic and molecular spectroscopy as well as interferometry experiments.
Similarly to dilaton-like DM~\cite{Arvanitaki:2014faa} or relaxion DM~\cite{Banerjee:2018xmn, Banerjee:2020kww}, atomic energy levels oscillate due to the axion DM background, leaving harmonic signals in the transition frequencies of stable frequency standards. 
By using existing experimental searches for scalar DM-SM interactions, we are able to place constraints on the axion-gluon interaction.  
Current constraints are generally weaker than existing constraints such as oscillating neutron EDM, but some of them already compete with astrophysical bounds e.g. SN1987A constraint. 
Future experiments based on nuclear clock transition or molecular vibrational excitation are expected to probe directly and model independently axion model in a region of parameter space that is currently unexplored.

\bigskip

{\it Note added}~~ While this work has been finalized, a related work~\cite{Davoudiasl:2022xms} appeared on arXiv, where the authors investigated the implications of the cosmological variation of $\bar\theta$.

\acknowledgements
We would like to thank Abhishek Banerjee, Fady Bishara, Joshua Eby, Zhen Liu, Oleksii Matsedonskyi, Ethan Neil, Pablo Quilez, Surjeet Rajendran, and Seokhoon Yun for useful discussions. 
We also thank Anson Hook for useful comments on the manuscript. 
We especially thank Eric Madge for pointing out numerical errors in the figure in the previous version of the manuscript.
The work of HK was supported by the Deutsche Forschungsgemeinschaft under Germany’s Excellence Strategy - EXC 2121 Quantum Universe - 390833306 and by the Munich Institute for Astro- and Particle Physics (MIAPP) which is funded by the Deutsche Forschungsgemeinschaft under Germany's Excellence Strategy – EXC-2094 – 390783311.
The work of GP is supported by grants from BSF-NSF, Friedrich Wilhelm Bessel research award, GIF, ISF, Minerva, SABRA - Yeda-Sela - WRC Program, the Estate of Emile Mimran, and the Maurice and Vivienne Wohl Endowment.

\bigskip

\appendix
\section{$\theta$-dependence of nuclear quantities}\label{app}
We detail the $\theta$-dependence of nuclear quantities.

The nucleon mass at the one-loop or chiral order ${\cal O}(p^3)$ is~\cite{Gasser:1987rb, Bernard:1992qa, Scherer:2012xha}
\bea
m_N(\theta) = m_0 - 4 c_1 m_\pi^2(\theta) - \frac{3 g_A^2 m_\pi^3(\theta)}{32\pi f_\pi^2} 
\eea
where $m_0$ is the nucleon mass in the chiral limit, $g_A = 1.27$ is the axial-vector coupling, and $c_1= - 1.1 \GeV^{-1}$ is a low energy constant~\cite{Hoferichter:2015tha}.
Therefore, we find
\bea
\frac{\partial \ln m_N}{\partial \ln m_\pi^2}  \simeq 0.06 ,
\label{mN_mpi}
\eea
where we have used $f_\pi =93\MeV$, and $m_N = 939\MeV$. 
Alternatively, we can write $\partial \ln m_N / \partial \ln m_\pi^2 = ( \sigma_{\pi N}/m_N)$. 
Current lattice computations predict $30\MeV \lesssim \sigma_{\pi N} \lesssim 70\MeV$ (see review~\cite{Aoki:2021kgd}), which translates into $0.03 \lesssim \partial \ln m_N / \partial \ln m_\pi^2 \lesssim 0.08$. 
We use Eq.~\eqref{mN_mpi} for the analysis in the main text.

The nucleon $g$-factor can also be computed from the chiral Lagrangian.
At the chiral order ${\cal O}(p^3)$, one finds~\cite{Scherer:2012xha}
\bea
g_p(\theta) &=& g_p^{(0)} - \frac{g_A^2 m_N m_\pi(\theta)}{4\pi f_\pi^2} ,
\\
g_n(\theta) &=& g_n^{(0)} + \frac{g_A^2 m_N m_\pi(\theta) }{4\pi f_\pi^2} .
\eea
Therefore, we find
\bea
\frac{\partial \ln g_p}{\partial \ln m_\pi^2}  &=& -\frac{1}{g_p} \frac{g_A^2 m_N m_\pi}{8 \pi f_\pi^2}
\simeq -0.17
\label{gp}
\\
\frac{\partial \ln g_n}{\partial \ln m_\pi^2} &=& + \frac{1}{g_n} \frac{g_A^2 m_N m_\pi}{8 \pi f_\pi^2}
\simeq -0.25
\label{gn}
\eea
where we have used $g_n = - 3.826$ and $g_p = 5.586$. 
The actual value might be smaller than the above values. 
One might extract isovector magnetic moment from form factors computed by lattice simulation~\cite{Djukanovic:2021cgp}, and the resulting dependence of $g$-factor on pion mass from such result is a factor two or three smaller than Eqs.~\eqref{gp}--\eqref{gn}. 
For the purpose of order of magnitude estimation, we use Eqs.~\eqref{gp}--\eqref{gn}. 

The nuclear $g$-factor can be written as a function of nucleon $g$-factor as well as spin expectation value of valence and core nucleons. 
Following Ref.~\cite{Flambaum:2006ip}, one finds
$$
\frac{\delta g}{g} 
=
\left[
K_n \frac{\partial \ln g_n}{\partial \ln m_\pi^2}
+ K_p \frac{\partial \ln g_p}{\partial \ln m_\pi^2}
- 0.17 K_b
\right]\frac{\delta m_\pi^2}{m_\pi^2}
$$
With values of $K_{n,p,b}$ given in~\cite{Flambaum:2006ip}, we find $\delta g /g = -0.02 (\delta m_\pi^2/m_\pi^2)$ for $^{87}$Rb and $\delta g /g = 0.01 ( \delta m_\pi^2 / m_\pi^2)$ for $^{133}$Cs.

The variation of nuclear clock transition in $^{229}$Th is estimated in Refs.~\cite{Flambaum:2007mj, Flambaum:2008hu}. 
\bea
&&
\frac{\delta f_{\rm Th}}{f_{\rm Th}} 
\nonumber \\
& \simeq &
1.3 \times 10^5 
\left( - 12 \frac{\delta m_N}{m_N} + 10 \frac{\delta m_\Delta}{m_\Delta}
+6 \frac{\delta m_\pi}{m_\pi}
- 43 \frac{\delta m_V}{m_V} 
\right)
\nonumber \\
& \simeq &
2 \times 10^5 \frac{\delta m_\pi^2}{m_\pi^2},
\eea
where $m_\Delta$ and $m_V$ are the masses of $\Delta$-baryon and vector meson. 
Among contributions from different mesons and hadrons, the pion contribution dominates all. 
It is straightforward to check that
\bea
\frac{\delta m_N}{m_N} 
&\simeq& 0.13 \frac{\delta m_\pi}{m_\pi}
\\
\frac{\delta m_\Delta}{m_\Delta} 
&\simeq& 0.03 \frac{\delta m_\pi}{m_\pi}
\\
\frac{\delta m_V}{m_V} 
&\simeq&  0.04 \frac{\delta m_\pi}{m_\pi}
\eea
where the variance of $\Delta$ baryon can be obtained directly from Eq.~(62) in~\cite{Bernard:2007zu} or from $\sigma_{\pi\Delta} = m_\pi^2 \partial m_\Delta / \partial m_\pi^2=20.6\MeV$. 
The variation of vector meson mass is obtained in~\cite{Flambaum:2007mj, Hanhart:2008mx}.

\bibliography{ref}
\end{document}